\documentclass[final,3p,times,twocolumn]{elsarticle}
\usepackage{amssymb}
\usepackage{amsmath}
\usepackage{url}
\usepackage{xcolor}
\journal{Knowledge-based Systems}

\begin{document}

\begin{frontmatter}

%% Title, authors and addresses

%% use the tnoteref command within \title for footnotes;
%% use the tnotetext command for theassociated footnote;
%% use the fnref command within \author or \affiliation for footnotes;
%% use the fntext command for theassociated footnote;
%% use the corref command within \author for corresponding author footnotes;
%% use the cortext command for theassociated footnote;
%% use the ead command for the email address,
%% and the form \ead[url] for the home page:
%% \title{Title\tnoteref{label1}}
%% \tnotetext[label1]{}
%% \author{Name\corref{cor1}\fnref{label2}}
%% \ead{email address}
%% \ead[url]{home page}
%% \fntext[label2]{}
%% \cortext[cor1]{}
%% \affiliation{organization={},
%%             addressline={},
%%             city={},
%%             postcode={},
%%             state={},
%%             country={}}
%% \fntext[label3]{}

\title{Performance of a GPU- and Time-Efficient Pseudo 3D Network for Magnetic Resonance Image Super-Resolution and Motion Artifact Reduction}

%% use optional labels to link authors explicitly to addresses:
%% \author[label1,label2]{}
%% \affiliation[label1]{organization={},
%%             addressline={},
%%             city={},
%%             postcode={},
%%             state={},
%%             country={}}
%%
%% \affiliation[label2]{organization={},
%%             addressline={},
%%             city={},
%%             postcode={},
%%             state={},
%%             country={}}

\author[UKHD]{Hao Li}
\author[Liu]{Jianan Liu}
\author[UKHD]{Marianne Schell}
\author[Huang]{Tao Huang}
\author[UKHD]{Arne Lauer}
\author[UKHD]{Katharina Schregel}
\author[UKHD]{Jessica Jesser}
\author[UKHD]{Dominik F Vollherbst}
\author[UKHD]{Martin Bendszus}
\author[UKHD]{Sabine Heiland}
\author[UKHD]{Tim Hilgenfeld\corref{cor1}} %% Author name
\cortext[cor1]{Corresponding author. Department of Neuroradiology, University Hospital Heidelberg, 69120 Heidelberg, Germany. \textit{E-mail address:} \textcolor{cyan}{\url{tim.hilgenfeld@med.uni-heidelberg.de}}}
%% Author affiliation
\affiliation[UKHD]{organization={Department of Neuroradiology, University Hospital Heidelberg},%Department and Organization
            addressline={Im Neuenheimer Feld 400}, 
            city={Heidelberg},
            postcode={69121}, 
            country={Germany}}
            
\affiliation[Liu]{organization={Momoni AI},%Department and Organization
%            addressline={}, 
            city={Gothenburg},
%            postcode={}, 
            country={Sweden}}
            
\affiliation[Huang]{organization={College of Science and Engineering, James Cook University},%Department and Organization
            addressline={14-88 McGregor Rd}, 
            city={Smithfield},
            postcode={QLD 4878}, 
            country={Australia}}

%% Abstract
\begin{abstract}
%% Text of abstract
Shortening acquisition time and reducing motion artifacts are the most critical challenges in magnetic resonance imaging (MRI). Deep learning-based image restoration has emerged as a promising solution capable of generating high-resolution and motion-artifact-free MRI images from low-resolution images acquired with shortened acquisition times or from motion-artifact-corrupted images. To facilitate clinical integration, a time- and GPU-efficient network with reliable accuracy is essential. In this study, we adopted a unified 2D deep learning framework for pseudo-3D MRI image super-resolution reconstruction (SRR) and motion artifact reduction (MAR). The optimal down-sampling factors to optimize the acquisition time in SRR were identified. Training for MAR was performed using publicly available in vivo data, employing a novel standardized method to induce motion artifacts of varying severity in a controlled way. The accuracy of the network was evaluated through a pixel-wise uncertainty map, and performance was benchmarked against state-of-the-art methods. The results demonstrated that the down-sampling factor of $1\times1\times2$ for $\times2$ acceleration and $2\times2\times2$ for $\times4$ acceleration was optimal. For SRR, the proposed TS-RCAN outperformed the 3D networks of mDCSRN and ReCNN, with an improvement of more than 0.01 in SSIM and 1.5 dB in PSNR while reducing GPU load by up to and inference time by up to 90\%. For MAR, TS-RCAN exceeded UNet's performance by up to 0.014 in SSIM and 1.48dB in PSNR. Additionally, TS-RCAN provided uncertainty information, which can be used to estimate the quality of the reconstructed images. TS-RCAN has potential use for SRR and MAR in the clinical setting.
\end{abstract}

%%Graphical abstract
%\begin{graphicalabstract}
%\includegraphics{grabs}
%\end{graphicalabstract}

%%Research highlights
%\begin{highlights}
%\item Research highlight 1
%\item Research highlight 2
%\end{highlights}

%% Keywords
\begin{keyword}
%% keywords here, in the form: keyword \sep keyword
%MRI, 3D Super Resolution, In-plane and Through-plane Down-sampling, Motion Artifact Generation, 3D Motion Artifact Reduction, Aleatoric and Epistemic Uncertainty, Deep Learning for Clinic.

Magnetic Resonance Imaging, 3D Super-Resolution Reconstruction, 3D Motion Artifact Reduction, Uncertainty for Image Quality Estimation, GPU and Inference Time Efficiency
%% PACS codes here, in the form: \PACS code \sep code

%% MSC codes here, in the form: \MSC code \sep code
%% or \MSC[2008] code \sep code (2000 is the default)

\end{keyword}

\end{frontmatter}

%% Add \usepackage{lineno} before \begin{document} and uncomment 
%% following line to enable line numbers
%% \linenumbers

%% main text
%%

%% Use \section commands to start a section
\section{Introduction}
\label{introduction}
%% Labels are used to cross-reference an item using \ref command.

Magnetic resonance imaging (MRI) is used in a wide range of medical applications to aid in precise diagnosis. However, diagnostic capabilities are frequently limited by spatial resolution and acquisition time constraints. The acquisition of high-resolution (HR) magnetic resonance images offers greater details but requires more scan time, resulting in a higher probability of motion artifacts (MA) that degrade image quality. 

Recent advances in deep learning offer a promising solution for enhancing image quality through techniques like super-resolution reconstruction (SRR) and motion artifact reduction (MAR). 

SRR is deeply influenced by deep learning technology. In such a data-driven approach, a great number of image pairs consisting of low-resolution (LR) and corresponding HR images are collected as training data, and the deep neural network is trained to extract pixel-wise features and generate the super-resolution (SR) image. Dong et al. \cite{1,2} first used this end-to-end learning-based method with a 2D convolutional neural network (CNN). Most available SRR studies in the field of radiology applied 2D network structures \cite{3,4}. However, computed tomography (CT) and MRI images typically contain information on 3D anatomical structures. Processing each image slice separately may lead to a mismatch in the adjacent slices in the reconstructed images. Consequently, 3D networks are desired to solve this problem due to their ability to extract 3D structural information \cite{5} and are proven to outperform 2D CNNs by a wide margin in MRI SRR \cite{6,7,8,9}. However, due to the dramatically increased number of operations of 3D CNNs, a substantial increase in GPU resources is needed. Even with that, much longer inference times are necessary, limiting the current application of 3D CNNs under clinical conditions. Although Chen et al. \cite{6,7} have demonstrated fast and GPU-efficient 3D networks, the gap of GPU consumption and inference time between 2D and 3D networks is still huge. 

The chosen down-sampling factors also have a crucial impact on reconstruction complexity, saving acquisition time and accuracy. In the studies of 2D networks, $2\times2\times1$ ($frequency-encoding(FE)\times phase-encoding(PE)\times slice-encoding(SL)$) and $4\times4\times1$ were commonly used \cite{10,11}. For 3D networks, HR images were down-sampled with varying scale factors, particularly with through-plane down-sampling, such as $1\times1\times2$, $1\times1\times4$ or $2\times2\times2$ \cite{6,12,13,14}. These down-sampling factors lead to various acceleration factors in acquisition time and different difficulties for SRR. To minimize GPU load and maximize acceleration of acquisition and SRR accuracy, a systematic analysis of down-sampling factors is necessary. However, this has not been addressed so far.

Motion artifacts are a very common problem in clinical MRI \cite{15}, often reducing diagnostic accuracy \cite{16,17}. Deep learning algorithms can reduce MA substantially \cite{18,19,20,21,22,23,24}. Although 3D UNet-based network was adopted in an early study for 3D MAR \cite{19}, 2D UNet-based networks were still predominantly used in nearly all follow-up studies for MAR \cite{18,20,21,22,23}. However, the disadvantage of 2D CNNs for SRR applications applies to MA reduction as well, limiting overall performance. 

Finally, concerns about the accuracy of deep learning-based image modifications, particularly in medical imaging, are valid. Incorrectly generated anatomical structures can lead to biased diagnosis or treatment decisions. To address these concerns, Tanno et al. \cite{25} and Qin et al. \cite{26} employed a method to predict the aleatoric uncertainty for the reconstructed image \cite{27} as auxiliary information to predict the accuracy. However, such a method fails to distinguish inherently different sources of uncertainty, including noise in training data or from the deep neural network due to a lack of knowledge when out-of-distribution (OOD) data are fed to the network \cite{28,29}. The former one is recognized as the aleatoric uncertainty, which cannot be eliminated. The latter can be reduced by extending the coverage of the training data. Epistemic uncertainty is the actual uncertainty that reflects the quality of the restored MR image, arising from the gap between the distributions of the training data and the test data which is commonly seen in real clinical environments. As CNN accuracy in the training setting may differ substantially compared to the clinical environment where ground truth (GT) images are not available, a quantitative estimation of CNN accuracy of restored MR images would be desirable. 

We hypothesized that the development of a unified, time- and GPU-efficient 2D CNN for 3D SRR and MAR with identical / superior performance compared to established 3D CNNs individually developed for SRR or MAR is possible. Furthermore, we propose that a pixel-wise uncertainty calculation is possible to estimate the accuracy in the absence of GT images. Therefore, the contributions of this manuscript are summarized as follows:
\begin{enumerate}
    \item[$\bullet$] We propose a unified 2D convolutional network for pseudo-3D SRR and MAR. The proposed network outperformed most state-of-the-art 3D networks in SRR while consuming highly reduced GPU resource and inference time.
    \item[$\bullet$] Optimal dowmsampling factors were identified to achieve the best SRR performance for specific acceleration factors.
    \item[$\bullet$] A standardized method for generating motion artifacts was proposed to create images with motion artifacts with different severeties in a controlled way.
    \item[$\bullet$] The proposed network was also capable of providing uncertainty maps that can be used to estimate the quality of the reconstructed images.
\end{enumerate}

\section{Method and Materials}
\label{method}

\subsection{MR image restoration Network}
\label{subsec2-1}

A thin slab residual channel attention network (TS-RCAN) was developed based on 2D RCAN \cite{30} for the restoration of 3D MR images. Figure \ref{fig1} shows the basic pipeline of the proposed network, which consists of several residual groups (RG) constructed with residual channel attention blocks (RCAB) and an up-sampling module if an in-plane upscaling is needed. The LR images with a single slice or a thin slab with several slices were fed to the network to generate the corresponding 2D/3D SR images.

\begin{figure*}[t]%% placement specifier
\centering%% For centre alignment of image.
\includegraphics[width=\textwidth]{fig1.eps}
\caption{Pipeline of thin-slab RCAN (TS-RCAN) used for MRI super resolution reconstruction and motion artifact reduction.}\label{fig1}
\end{figure*}

The 2D network is intrinsically capable of processing 3D datasets with its channel dimension. Therefore, the network was applied in multichannel input and multichannel output mode, with the third dimension of the 3D image patch placed on the channel dimension. When the first convolutional layer of the network has the number of input channels as $M$ (we used $M\le5$ in our experiments) and the size of the input image patches $M\times H\times W$ ($M$: number of channels, $H$: height of the matrix, $W$: width of the matrix). The size of the convolution kernel of the first layer is $B\times M\times H\times W$ ($B$: batch size), which works equally well with a 3D convolution, whose size is $B\times1\times M\times H\times W$ with non-padding. In this step, the features of multiple input slices were extracted and compressed into a single channel feature map with a certain number of different filters according to the number of channels in the hidden layers. Then, these extracted features are learned to reconstruct the high-quality images in the hidden layers. In the last convolutional layer, the number of output channels is equal to the expected slice number of the target patch. We used a thinner slab of input patches than the 3D patches for conventional 3D networks, since the convolution kernels did not stride along the third dimension in the 2D network. 

In addition, conventional network-based \cite{31} and data-based ensemble methods \cite{32} require more training time or operation complexity. Therefore, a simple and effective approach to self-ensemble was used to further improve performance. The thin-slab patches consisted of several slices, each of which appeared in different places in different patches and was processed by the network differently, resulting in multiple outputs for the same slice.

\subsection{Down-sampling factors for super-resolution}
\label{subsec2-2}

Regarding the generation of synthetic LR image generation, K-space truncation is recently recognized as a way to mimic real LR image acquisition in MRI \cite{11}. Therefore, we also adopted this method to generate synthetic LR images in our study. 

HR images were transformed to K-spaces using 3D Fast Fourier Transformation (FFT), and the K-spaces were truncated in three dimensions according to the scale factors and only the central region of the K-spaces remained. Then the truncated K-spaces were transformed to LR images using 3D inverse FFT (iFFT). Finally, the intensities of the voxels for both HR and LR images were scaled to the range of 0 to 1. Figure \ref{fig2}(A) shows an example of 3D LR MR image generation.

\begin{figure*}[t]%% placement specifier
\centering%% For centre alignment of image.
\includegraphics[width=325pt]{fig2.eps}
\caption{Retrospective generation of 3D low resolution and MA-corrupted MR images. (A): Generation of 3D low resolution images with scale factor of $2\times2\times2$ and patch cropping. (B): Image-based generation of MA in MR images. (C): Scheme of motion pattern employed in our study.}\label{fig2}
\end{figure*}

The acquisition time of 3D MRI highly depends on phase-encoding steps in the PE and SL directions, thus only down-sampling in PE and SL directions shorten the acquisition time. However, in regular MRI measurements, the resolutions in the PE and FE directions always change simultaneously with the same scale factor to maintain the isotropic in-plane resolution, resulting in an unnecessary loss of K-space in the FE direction and greater difficulty in restoring the HR image. Meanwhile, the slice thickness is independent of the in-plane resolution and can be down-sampled with greater flexibility. More specifically, when a $\times2$ acceleration is expected, there are two down-sampling options, which are $2\times2\times1$ and $1\times1\times2$. With the $2\times2\times1$ down-sampling, 75\% of the K-space is dropped, while with $1\times1\times2$ only 50\% of the K-space is dropped. 

However, the difficulty in restoring the HR image is not solely determined by the ratio of K-space truncation. The more low-frequency components of the K-space are dropped, the more difficult it is to restore the HR image. In this study, we investigated multiple down-sampling factors and their SRR performances. Based on the regular MRI measurement process, we tested the SRR with the down-sampling factors of $2\times2\times1$ and $1\times1\times2$ for $\times2$ acceleration, and $4\times4\times1$, $2\times2\times2$, and $1\times1\times4$ for $\times4$ acceleration.

After down-sampling, the HR and LR images were cropped into patches with smaller sizes to save computation resources.  HR images were cropped into $128\times128$ patches with 32 voxels overlapped between neighboring patches to avoid artifacts on the edges of the patches. LR images were cropped into $64\times64$ patches with 16 voxels overlapped for the in-plane scale factor of 2 and $32\times32$ patches with 8 voxels overlapped for the in-plane scale factor of 4. Each LR patch contains 1/3/5 successive slices from the LR image with $n-1$ slices overlapped between neighboring patches, and the number of slices for the HR patch is 1/3/5 times the through-plane scale factor. For the 3D networks used for comparison, the LR images were interpolated to the same matrix size as the HR images \cite{6,7,8,33}, since the 3D networks do not have an up-sampling module. Both HR and LR images were cropped into $64\times64\times64$ patches with 32 voxels overlapped between neighboring patches.

\subsection{Motion pattern and motion artifact quantification}
\label{subsec2-3}

The method of splicing lines from multiple K-spaces was used to simulate the generation of real MA in MR images. As shown in Figure \ref{fig2}(B), a group of images was generated from the original image volume by rotating it in specific directions with certain degrees. The original image and generated images were then transformed to K-space using FFT, and K-space segments of the original image were replaced by the segments from the generated images' K-spaces based on the predefined pattern. The images for the motion artifact correction task were not cropped, so the axial sizes of the input MA and GT images were $320\times256$.

Regarding movement patterns, previous studies used random movement to retrospectively generate MA \cite{22,23,24}, making the severity of MA uncontrollable and irreproducible. Therefore, we used simplified periodic patterns of motion, which were a head rotation of 5 degrees to the left and right as the in-plane movement and a head nodding of 5 degrees as the through-plane movement, and the severity was controlled by the duration and frequency of motion. The scheme of the motion pattern is illustrated in Figure \ref{fig2}(C). We used echo-group (EG) as a unit of the minimal time period in which a certain number of successive echoes were acquired (which can also be considered as the TR for sequences from the turbo-spin-echo family), and the duration of any action must be integer multiples of the EG. Therefore, the whole process of patient movement was set as below:
\begin{enumerate}
\item[$1)$] At $t=0$, the patient stayed in the original position and stayed for $T_S$;
\item[$2)$] from $t=T_S$ to $t=T_S+2EG$, the patient's head rotated to the left for 5 degrees;
\item[$3)$] from $t=T_S+2EG$ to $t=T_S+7EG$, the patient's head stayed at the position of 5 degrees to the left;
\item[$4)$] from $t=T_S+7EG$ to $t=T_S+9EG$, the patient's head rotated back to the starting position;
\item[$5)$] from $t=T_S+9EG$ to $t=2T_S+9EG$, the patient's head stayed in the starting position;
\item[$6)$] from $t=2T_S+9EG$ to $t=2T_S+18EG$, the patient's head rotated to the right and returned to the starting position following the same process of steps 2 to 4;
\item[$7)$] from $t=2T_S+18EG$ to $t=3T_S+18EG$, the patient's head stayed in the starting position.
\end{enumerate}

The process from steps 2 to 7 was repeated until the whole K-space was acquired, and head nodding was performed together with the head rotating. The severity of MA was controlled by $T_s$ and $EG$. 

In our study, $T_s$ was set to $9EG$,$18EG$,$36EG$ and $72EG$, leading to a corrupted K-space ratio of 50\%,33.3\%,20\% and 11.1\%. $1EG$ consists of 80 echoes, and a centric trajectory was selected to fill the K-space in our study. The image quality metrics values of the images with different severities of MA follow a linear tendency as shown in Table 1 in the result section.

\subsection{Uncertainty}
\label{subsec2-4}

In previous studies, the pixel-wise aleatoric uncertainty can be estimated by incorporating Gaussian negative log-likelihood (NLL) loss into neural network \cite{34} and have been applied in the SRR of MRI \cite{25,26}. This uncertainty only represents the uncertainty of the training data, which can be mitigated by enlarging the training dataset. Therefore, it is not the main issue of medical image restoration in clinical practice. In contrast, the inevitable issue of OOD data (i.e., images acquired from different patients or from a different scanner may have different distributions in a real clinic environment, even with the same protocols), which can be represented by epistemic uncertainty, is more crucial \cite{27}. In our study, both pixel-wise aleatoric uncertainty and epistemic uncertainty are estimated using evidential regression \cite{29}. Evidential deep learning formulates learning as an evidence acquisition process. Every training example adds support to a learned higher-order evidential distribution. Sampling from this distribution yields instances of lower-order likelihood functions from which the data were drawn. Instead of placing priors on network weights, as is done in Bayesian neural networks, evidential approaches place priors directly over the likelihood function. By training a neural network to output the hyperparameters of the higher-order evidential distribution, a grounded representation of both epistemic and aleatoric uncertainty can then be learned without the need for sampling. Amini et al. \cite{29} proposed the method to estimate a posterior distribution $q(\mu,\sigma^2 )$, and the approximation to the posterior distribution takes the form of the Gaussian conjugate prior, the Normal Inverse-Gamma (NIG) distribution $p(\mu,\sigma^2\vert\gamma,\nu,\alpha,\beta)$. Afterwards, the prediction and uncertainties can be calculated as:
\begin{equation}
Prediction: E[\mu]=\gamma
\end{equation}
\begin{equation}
Aleatoric: E[\sigma^2]=\frac{\beta}{\alpha-1}
\end{equation}
\begin{equation}
Epistemic: Var[\mu]=\frac{\beta}{\nu(\alpha-1)}
\end{equation}

Furthermore, we investigated the correlations of epistemic uncertainty with the structure similarity index (SSIM) \cite{35} and peak signal-to-noise ratio (PSNR) values of the reconstructed images. Linear regression and exponential regression were performed to estimate the correlation of epistemic uncertainty with SSIM and PSNR for the test datasets, respectively. Then, the obtained regression curves can be used to estimate the SSIM and PSNR of the reconstructed images when the GT is not available.

\begin{figure*}[t]%% placement specifier
\centering%% For centre alignment of image.
\includegraphics[width=\textwidth]{fig3.eps}
\caption{Dependency of SRR performance on down-sampling factors. The scale factors are grouped according to the acceleration factor. The $1\times1\times2$ and $2\times2\times2$ down-sampling with $M\ge3$ achieved the highest mean values of SSIM/PSNR values for $\times2$ and $\times4$ acceleration respectively. With $\times2$ acceleration, $1\times1\times2$ down-sampling with $M=5$ and self-ensemble significantly outperformed all cases of $2\times2\times1$ in PSNR and SSIM. With $\times4$ acceleration, $2\times2\times2$ down-sampling with $M\ge3$ and self-ensemble significantly outperformed $4\times4\times1$ in SSIM. (A)/(C): PSNR and SSIM of super-resolution images with $\times2$ acceleration ($2\times2\times1$ and $1\times1\times2$); (B)/(D): PSNR and SSIM of super-resolution images with $\times4$ acceleration ($4\times4\times1$, $1\times1\times4$ and $2\times2\times2$). ‘+’ represents the SRR with self-ensemble. Significant difference is indicated with $*p<0.05$ and $**p<0.0005$.}\label{fig3}
\end{figure*}

\subsection{Loss functions}
\label{subsec2-5}

In previous studies, different types of loss function were used to train neural networks for specific feature refinements. In this study, pixel-wise Charbonnier loss \cite{36} was used as a differentiable L1 loss to avoid a strong smoothing effect \cite{37}: 
\begin{equation}
L_{Char}=\frac{1}{N}\sum_{i=1}^N \sqrt{(HR_i-SR_i)^2+\epsilon}
\end{equation}
where $\epsilon$ is assigned as $10^{-4}$.

Furthermore, in recent studies, SSIM loss has been widely used to drive the network to reconstruct high-quality images toward better structural similarity to GT images \cite{11}. In this study, we utilized the L1 loss of the square of the SSIM value to enhance the weight of SSIM loss:
\begin{equation}
L_{SSIM}=\frac{1}{N}\sum_{i=1}^N (1-SSIM(SR_i,HR_i)^2)
\end{equation}

In our study, we utilized the weighted sum of Charbonnier loss and SSIM loss for image restoration:
\begin{equation}
Loss=L_{Char}+w_1L_{SSIM}
\end{equation}
where $w_1=0.5$ in our study for the best performance.

Besides, the Normal-inverse-Gamma loss \cite{29} was employed when the uncertainty map was demanded:
\begin{equation}
L_{NIG}=L_{NLL}+\lambda L_{Reg}
\end{equation}
where
\begin{equation}
\begin{split}
L_{NLL}&=\frac{1}{2}log(\frac{\pi}{\nu})-\alpha log(\Omega)+log(\frac{\Gamma(\alpha)}{\Gamma(\alpha+\frac{1}{2})})\\
&+(\alpha+\frac{1}{2})log((y-\gamma)^2\nu+\Omega)
\end{split}
\end{equation}
\begin{equation}
\Omega=2\beta(1+\nu)
\end{equation}
\begin{equation}
L_{Reg}=\lvert y-\gamma\rvert(2\nu+\alpha)
\end{equation}
$\alpha$, $\beta$, $\gamma$ and $\nu$ were the outputs of the network. $y$ was the GT.

To summarize, we utilized the weighted sum of Charbonnier loss and SSIM loss for image restoration, and NIG loss when evidential regression was employed:
\begin{equation}
Loss=L_{Char}+w_1L_{SSIM}+w_2L_{NIG}
\end{equation}
where 0.5 and 1 were selected as the value of $w_1$ and $w_2$ in our study for the best performance.

\begin{figure}[t]%% placement specifier
\centering%% For centre alignment of image.
\includegraphics[width=222pt]{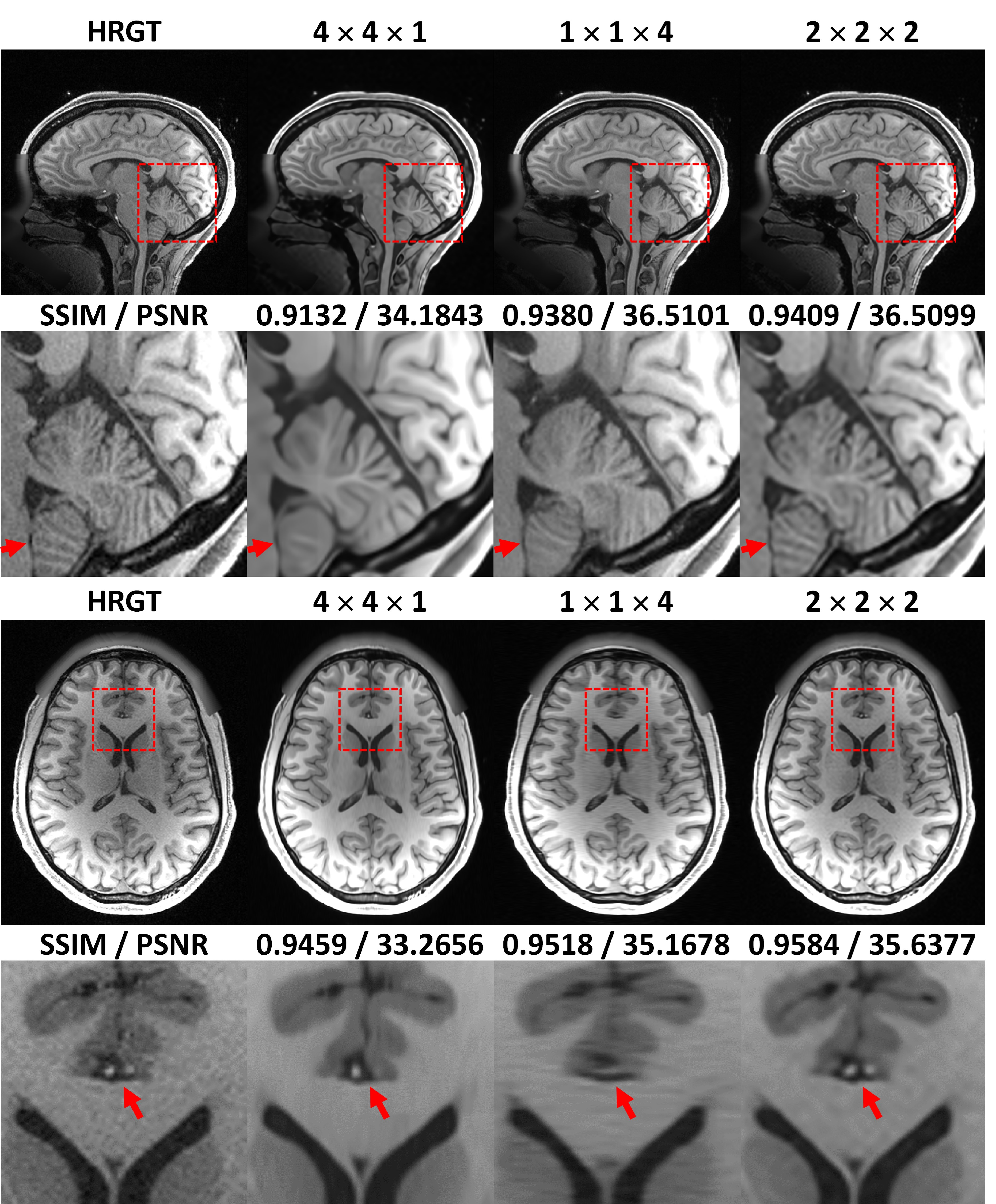}
\caption{Qualitative comparison of 3D SRR with different down-sampling factors of $\times4$ acceleration. Note the loss of anatomical details in the cerebellar grey-white-matter differentiability in the sagittal image (arrow) of the $4\times4\times1$ down-sampling strategy, as well as the loss of one/both anterior cerebral arteries (arrows) in the axial images of the $4\times4\times1$ and $1\times1\times4$ down-sampling factor. Best qualitative results were achieved using the $2\times2\times2$ down-sampling factor.}\label{fig4}
\end{figure}

\subsection{Datasets}
\label{subsec2-6}

In this study, we used the T1w images from the Human Connectome Project (HCP) dataset \cite{38}, which consists of multi-contrast images from 1113 patients. The T1w images were acquired in the sagittal direction with 3D MPRAGE on the Siemens 3T PRISMA platform. The size of the matrix was $320\times320\times256$, with an isotropic resolution of 0.7 mm. In our experiments, we randomly selected 80/10/10 patients from the HCP patient cohort for training/validation/test groups. The training datasets were used to train neural networks, the validation datasets to monitor the neural networks' performance during training, and the test datasets to evaluate neural networks after training. There were no shared datasets among the three groups. Furthermore, to verify the quantified correlation between uncertainty and SSIM/PSNR, we used another 40 patients from the HCP dataset as the accuracy prediction dataset, which were isolated from the previously mentioned training/validation/test groups.

\subsection{Implementation details}
\label{subsec2-7}

In terms of TS-RCAN implementation, the network was formed by 5 RGs, with 5 RCABs in each RG. The convolution layers in shallow feature extraction and the residual in residual structure had 64 filters. The networks were trained on a workstation equipped with two Nvidia Quadro A6000 graphics cards and PyTorch 1.9. In each training batch, eight LR patches were randomly extracted as inputs. We trained TS-RCAN for 50 epochs using the ADAM optimizer with $\beta_1=0.9$,$\beta_2=0.999$, and $\upsilon=10^{-8}$, and a Cosine-decay learning rate was applied with a learning rate of $10^{-4}$ to $10^{-8}$. We employed PSNR and SSIM \cite{35} to evaluate the quality of the reconstructed images. The normal distribution of the data was examined using the Shapiro-Wilk test. One-way analysis of variance (ANOVA) test with post hoc Tukey test and Kruskal-Wallis test with post hoc Dunn’s test were applied to test the statistical difference for normal distributed and nonnormal distributed data, respectively. Mann-Whitney U tests were performed in the head-to-head comparison to detect statistical differences between the results of the state-of-the-art methods and our method. All data processing was performed using MATLAB and statistical analysis was performed using GraphPad Prism.

\begin{figure*}%% placement specifier
\centering%% For centre alignment of image.
\includegraphics[width=\textwidth]{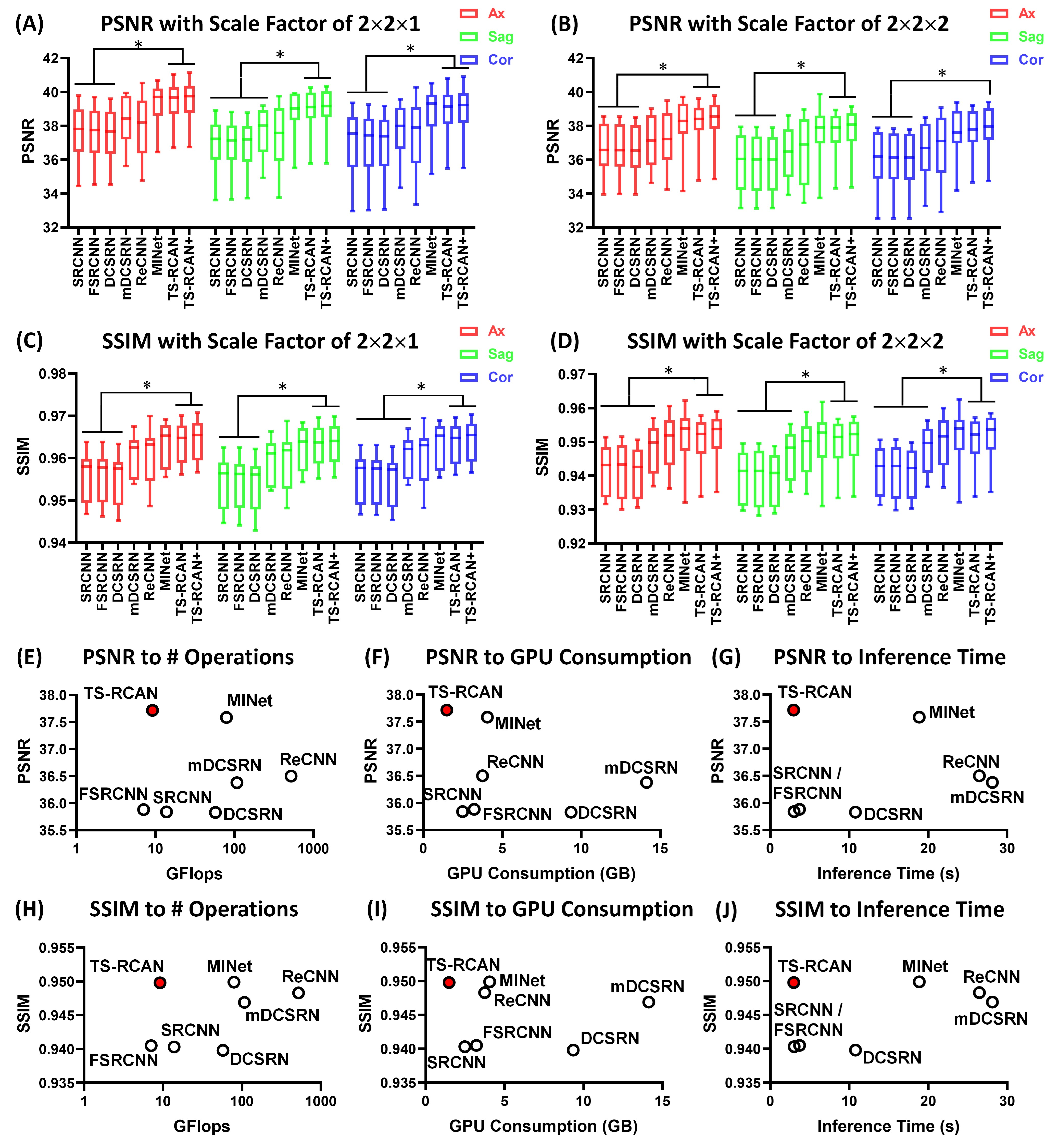}
\caption{Comparison of TS-RCAN with other state-of-the-art 3D networks. (A)-(D): comparison of metrics with two scale factors ($2\times2\times1$ and $2\times2\times2$), ‘+’ represents the SRR with self-ensemble. TS-RCAN achieved comparable performance with MINet and outperformed the other networks in SSIM and PSNR. Significant difference is indicated with $*p<0.05$ and $**p<0.0005$. (E)-(J): Comparison of the number of operations, GPU consumption and inference time to PSNR and SSIM with scale factor of $2\times2\times2$. TS-RCAN achieved top performance by consuming the minimal computation resources and inference time.}\label{fig5}
\end{figure*}

\begin{figure*}[t]%% placement specifier
\centering%% For centre alignment of image.
\includegraphics[width=\textwidth]{fig6.eps}
\caption{Qualitative comparison of 3D MRI SRR to the state-of-the-art methods with the scale factor of $2\times2\times2$. The reconstructed SRR images of SRCNN, FSRCNN and DCSRN were blurry in both sagittal and axial view, resulting in a nearly complete absence of gray matter identification in the hand knob area (arrow). In contrast, ReCNN, MINet and TS-RCAN showed comparable performance with lower error and enhanced distinguishability of small anatomical structures (arrow).}\label{fig6}
\end{figure*}

\section{Experiments and Results}
\label{results}

\subsection{Dependency of SRR network performance on down-sampling factors}
\label{subsec3-1}

The mean SSIM and PSNR for the SRR of the factors with only in-plane ($2\times2\times1$ and $4\times4\times1$) down-sampling with 3-slice input ($M=3$) outperformed the SRR with $M=1$ by up to 0.004 / 0.3 dB (p-values in the range of $>0.99$). After reaching their peaks at $M=3$, the mean SSIM and PSNR dropped slightly with 5-slice input ($M=5$) (Figure \ref{fig3}). For factors with through-plane down-sampling ($1\times1\times2$, $1\times1\times4$ and $2\times2\times2$), the mean SSIM and PSNR of SR images reconstructed with $M=3$ soared from $M=1$ with significant difference in most cases, as shown in Figure \ref{fig3} (p-values in the range of $<0.0001$ to 0.03). The mean SSIM and PSNR increased again with $M=5$ for factors with through-plane down-sampling, although no significant differences were detected (p-values in the range of $>0.99$). In addition, self-ensemble improved the performance of the 3D mode in the SRR for all down-sampling factors.

Furthermore, SRR performance was compared between different down-sampling factors with the same acceleration factors. With $\times2$ acceleration, the mean PSNR value of $1\times1\times2$ down-sampling with $M=5$ and self-ensemble was significantly over 2.2 dB higher than those of $2\times2\times1$ with different configurations in three directions (p-value range: 0.01 to 0.04). The mean SSIM values of $1\times1\times2$ down-sampling with multi-slice input ($M=3$  and 5) were significantly higher than those of $2\times2\times1$ by more than 0.01 with different configurations ($M=1$, 3 and 5) in three directions (p-value range: $<0.0001$ to 0.01). 

Moreover, with $\times4$ acceleration, the $2\times2\times2$ down-sampling has the highest mean values of SSIM/PSNR, which were over 0.003/0.3 dB higher than $1\times1\times4$ and 0.02/1.9 dB higher than $4\times4\times1$ in three directions, significant differences were detected in some cases, as shown in Figure \ref{fig3} (p-value range: 0.01 to 0.05). 

The higher SSIM/PSNR of $2\times2\times2$ down-sampling also resulted in a more accurate display of fine anatomical details (Figure \ref{fig4}). The reconstructed image of $4\times4\times1$ loses numerous small anatomical structures in the sagittal direction and $1\times1\times4$ in the axial direction, while the image reconstructed from $2\times2\times2$ is less blurry than the previous two, with the majority of small anatomical structures retained.

\subsection{SRR - head-to-head performance of various CNN}
\label{subsec3-2}

Previously published state-of-the-art networks, including 3D SRCNN \cite{1}, 3D FSRCNN \cite{2}, DCSRN \cite{6}, mDCSRN \cite{7}, ReCNN \cite{33} and MINet \cite{39} were implemented for performance comparison with TS-RCAN. All networks were trained with identical settings (i.e. epochs, learning rate, and optimizer). 

For SRR with scale factors of $2\times2\times1$ and $2\times2\times2$, TS-RCAN outperformed SRCNN, FSRCNN, and DCSRN in SSIM and PSNR with significant differences (p-value range: 0.004 to 0.045; Figure \ref{fig5}(A)-(D)). Although no significant differences were observed, the mean SSIM/PSNR of TS-RCAN were higher than mDCSRN and ReCNN, and was comparable to MINet which used a similar backbone as TS-RCAN.

\begin{figure*}[t]%% placement specifier
\centering%% For centre alignment of image.
\includegraphics[width=\textwidth]{fig7.eps}
\caption{Performance comparison of UNet and TS-RCAN networks for restoration of image quality in most motion artifact degraded image groups ($T_s=9EG$ and $18EG$). TS-RCAN with $M=3$ significantly improved image quality in all cases, which was not the case for UNet and TS-RCAN with $M=1$. Moreover, TS-RCAN outperformed UNet in SSIM/PSNR improvement in all directions / severities of motion artifacts. (A)/(C): PSNR and SSIM of MA-corrupted and reduced images with 5 degrees in-plane rotation; (B)/(D): PSNR and SSIM of MA-corrupted and reduced images with 5 degrees in-plane rotation and 5 degrees through-plane rotation. ‘+’ represents the MAR with self-ensemble. Significant differences are indicated with $*p<0.05$ and $**p<0.0005$.}\label{fig7}
\end{figure*}

When comparing the computation resources for $2\times2\times2$ SRR, the number of operations of TS-RCAN was comparable to 3D SRCNN and 3D FSRCNN, and significantly lower than the other networks. The GPU consumption and inference time were lower than all other networks (p-value range: $<0.0001$). Compared to ReCNN / MINet, whose SRR performance was closest to TS-RCAN, TS-RCAN consumed 60.4\% / 63.5\% lower VRAM. The mean inference time of TS-RCAN for processing the whole image volume of one patient was 11.2\% / 15.7\% of ReCNN / MINet, respectively.

In the qualitative comparison of 3D MRI SRR methods with a scale factor of $2\times2\times2$ in Figure 6, the SR images reconstructed by SRCNN, FSRCNN, and DCSRN appeared blurry in both the sagittal and axial views. This resulted in an almost complete loss of gray matter identification in the hand knob region (indicated by the arrow in the lower row). In contrast, ReCNN, MINet, and TS-RCAN demonstrated significantly better performance, with reduced error and improved distinguishability of small anatomical structures (arrow in both upper and lower rows), offering enhanced clarity and anatomical detail.

\begin{figure*}[t]%% placement specifier
\centering%% For centre alignment of image.
\includegraphics[width=\textwidth]{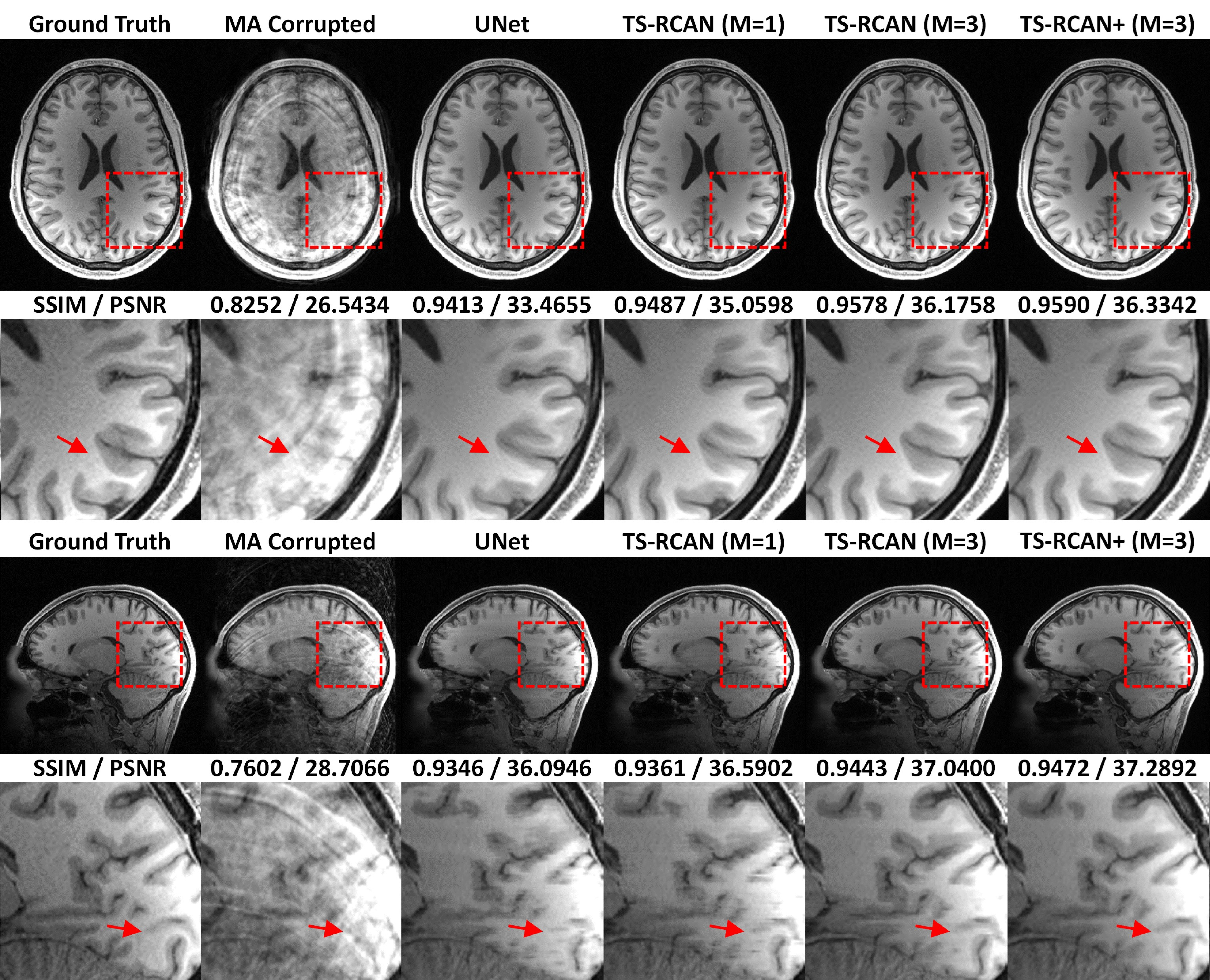}
\caption{Qualitative comparison on axial and sagittal planes for motion artifact reduction in group with most severe motion artifacts ($T_S=9EG$, and 5 degrees in-plane and through-plane rotation). }\label{fig8}
\end{figure*}

\subsection{Motion artifact reduction – performance of TS-RCAN vs. UNet}
\label{subsec3-3}

The MA-corrupted images generated by the proposed periodic MA generation algorithm revealed a mean drop in SSIM and PSNR by 0.049 to 0.074 and 2.5 to 3.3 dB when motion increased (by reducing $T_s$ by 50\% each time). The through-plane rotation also resulted in additional decrements of 0.003 to 0.012 in SSIM and 0.4 to 0.5 dB in PSNR compared to only in-plane rotation.

For all evaluated motion artifact severities, TS-RCAN ($M=3$, both with and without self-ensemble) significantly improved the mean SSIM / PSNR of MA reduced images compared to the noncorrected data sets (increment of mean SSIM / PSNR range: 0.18 to 0.20 / 6.33 to 8.77 dB; p-value range: 0.005 to 0.013 for axial direction of $T_s=9EG$ in Figure \ref{fig7}(C), and $\le0.001$ for the other cases). This improvement was not observed in the UNet results. In direct comparison to 2D UNet, TS-RCAN showed improved performance: the mean SSIM / PSNR increased by over 0.004 / 0.5 dB with $M=1$, and up to 0.014 / 1.48 dB with $M=3$ for $T_s=9EG$, although no significant differences were detected (p-value range: 0.22 to $>0.99$; Figure \ref{fig7}). The difference between UNet and TS-RCAN was most pronounced in datasets with the most severe MA.

\begin{figure*}[t]%% placement specifier
\centering%% For centre alignment of image.
\includegraphics[width=\textwidth]{fig9.eps}
\caption{Quantitative and qualitative analysis of uncertainty. (A): the GT image; (B): the SR image; (C): the SSIM map; (D): the absolute error map between GT and SR images; (E): aleatoric uncertainty map rescaled to the range of 0 to 1: showing that aleatoric uncertainty pervades the entire image volume, particularly the background region, which was dominated by random noise; (F) epistemic uncertainty map rescaled to the range of 0 to 1: highlighting the anatomical regions. High values in the epistemic uncertainty map corresponds to low/high values in the SSIM/ absolute error map; (G)/(H): Regression of SSIN/PSNR to epistemic uncertainty: the linear regression on SSIM and exponential regression on PSNR of accuracy test data (blue triangles) achieved $R^2$ values above 0.8, indicating the regressions fit the data well. 93.7\% / 95.5\% of accuracy prediction data (green crosses) located in the 95\% predict intervals of SSIM/PSNR.}\label{fig9}
\end{figure*}

In the qualitative comparison, the results of the quantitative analysis can be observed, with improved image quality of all networks compared to motion-corrupted source data, and the highest image quality achieved by TS-RCAN + with $M=3$ (Figure \ref{fig8}). In the axial plane, the UNet-corrected image contained several incorrectly restored anatomical structures, while the quality of the restored image using TS-RCAN with $M=1$ was improved with reduced errors (red arrow). With $M>1$, TS-RCAN provided significantly improved image quality, with most anatomical structures well represented. The difference between the networks is even greater in the sagittal plane. Due to the lack of through-slice information, the UNet-corrected image contained a severe through-slice mismatch, which was slightly reduced by TS-RCAN with $M=1$ and significantly reduced when $M=3$ (red arrow).

\subsection{Uncertainty evaluation}
\label{subsec3-4}

By investigating the mean epistemic uncertainty of each image slice and the SSIM/PSNR value, we observed strong correlations between them. As shown in Figure \ref{fig9}(G)(H), we performed a linear regression analysis on the mean epistemic and SSIM values using the 10 patients from the test group (blue triangles; accuracy test dataset), and obtained a regression equation (the red solid line) with a 95\% prediction interval (the region between the red dashed lines). Due to the logarithmic function of PSNR, an exponential regression was performed on mean epistemic and PSNR values, and a regression equation with a 95\% prediction interval was also obtained. The $R^2$ values of the regressions were greater than 0.8, indicating that the regressions fit the data well. Then we involved 40 additional datasets, which were isolated from the training/validation/test groups, to validate the accuracy of the regressions in predicting the SSIM and PSNR from the mean epistemic uncertainty values. The additional 40 datasets are represented as green crosses (accuracy prediction dataset) in Figure \ref{fig9}(G)(H) with 93.7\%/95.5\% of the data located in the prediction intervals, indicating that the correlations between mean epistemic uncertainty and SSIM/PSNR closely follow the predicted distribution.

\section{Discussion}
\label{discussion}

A GPU-efficient, unified 2D deep neural network (TS-RCAN) was proposed for pseudo-3D MRI super-resolution reconstruction and motion artifact reduction, resulting in superior image quality. The performance of TS-RCAN was equal to or better compared to previous SRR or MA reduction CNNs. In contrast to previously published 3D convolutional neural networks, the advanced performance of TS-RCAN for SRR was not dependent on high GPU workload and long inference time. Furthermore, the evaluation of the pixel-wise uncertainty of the TS-RCAN network yielded robust accuracy values. These results are of clinical importance as acquisition time and impact of motion are critical factors in clinical MRI protocols, especially in high-resolution 3D sequences. Consequently, both features of this network combined with its high accuracy could improve diagnostic accuracy of clinical MRI protocols across various MRI applications. 

This study has several strengths. First, this study comprehensively analyzed the impact of various down-sampling factors and identified when $M>1$ with $1\times1\times2$ for $\times2$ acceleration and $2\times2\times2$ for $\times4$ acceleration results in the best SRR performance. Self-ensemble further improved the performance of the network without any additional operations. Therefore, these down-sampling factors are recommended in deep learning-based SRR in clinical environments. Second, a combined CNN for SRR and MA reduction was developed. Third, the performance of the novel CNN was directly compared with existing CNNs in a head-to-head comparison. Regarding SRR, we confirmed the results of Chen et al. \cite{6,7}, Pham et al. \cite{8}, and Koktzoglou et al. \cite{9}, who reported that 3D SRR outperformed 2D SRR. TS-RCAN significantly outperformed 3D SRCNN \cite{1}, 3D FSCNN \cite{2} and DCSRN \cite{6} in 3D SRR by over 0.01 in SSIM and 1.9 dB in PSNR with scale factor of $1\times1\times2$, and over 0.01 in SSIM and 1.5 dB with scale factor of $2\times2\times2$. TS-RCAN also outperformed mDCSRN \cite{7} and ReCNN \cite{33}, although no significant differences were detected. TS-RCAN showed performance comparable to that of MINet \cite{39}, which had the same backbone network. 

Regarding MAR, TS-RCAN significantly outperformed than 2D UNet \cite{20,21}, particularly with the improved through-slice agreement. Instead of the random motion patterns used in previous studies for retrospective MA generation \cite{22,23,24}, our method adopted a predefined motion pattern with adjustable duration and frequency of movement, resulting in a controllable and quantifiable severity. MA-corrupted images with linearly increased SSIM and PSNR values were obtained by adjusting the frequency of movement in our experiments. The motion pattern can be modified based on any specific scenario.

Several methods using 2D neural networks to reconstruct 3D MR images were proposed in previous studies, but each had drawbacks. Du et al. proposed to interpolate the LR image to the expected size before feeding the 2D network slice by slice \cite{40}. However, this method still reconstructed 2D images without fusing the features from multiple slices in an image volume. Zhang et al. proposed to use 2D network to reconstruct the in-plane SR images in three orthogonal planes, and use another 2D network to combine the three groups of reconstructed SR images \cite{41,42}. Sood and Rusu et al. proposed a similar method, reconstructing 2D slices in two orthogonal planes and then rebuilding the 3D image volumes \cite{43}. These methods highly increased the complexity and demand several times of the computation resources of single 2D networks. Georgescu et al. proposed to process the 3D image volume with two networks progressively \cite{44}. A 2D network was used to process the images slice by slice for in-plane SRR, and a 3D network was used for through-plane SRR. In contrast, we propose using a single 2D network with single step of processing to reconstruct 3D image slabs. 

Regarding the consumption of GPU resource and inference time, TS-RCAN consumed comparable GPU resource and inference time with 3D SRCNN \cite{1} and 3D FSRCNN \cite{2}, down to 15.8\% / 10.5\% / 39.6\% / 36.5\% of GPU resources and less than 27.4\% / 10.2\% / 11.2\% / 15.7\% of inference time compared to DCSRN \cite{6} / mDCSRN \cite{7} / ReCNN \cite{33} / MINet \cite{39}. Thus, it can be easily deployed on any consumer GPU and reconstruct the image very timely. 

Finally, we adopted evidential regression learning to estimate uncertainty maps simultaneously with image restoration. Qin et al. have demonstrated the uncertainty map to predict the accuracy of reconstructed super-resolution images \cite{26}, but their method could not distinguish different sources of uncertainty. In contrast, the method adopted in this manuscript separated the aleatoric and epistemic uncertainties. Experimental results revealed that the aleatoric uncertainty highly depended on the noise from the training data. In addition, the epistemic uncertainty map corresponded to the absolute error map and the SSIM map between the ground truth and the restored image. We also investigated the correlation of epistemic uncertainty with SSIM and PSNR values. Our experiments revealed that the mean epistemic uncertainty of each image slice appeared to be linearly related to the SSIM and exponentially related to the PSNR. Thus, even when ground truth is not available in clinical settings, the SSIM and PSNR values can be predicted using the regression equations. Furthermore, the epistemic uncertainty map could also help doctors identify which regions of the reconstructed SR images are more reliable and avoid misguided diagnosis or treatments caused by incorrectly generated content. 

We acknowledge several limitations. First, images with motion artifacts were generated synthetically and may not reflect more irregular motion patterns that can be encountered under clinical circumstances, limiting the transferability of the results. Second, although very promising results of network performance and accuracy were observed, more studies in patients are necessary for a final assessment of usefulness.

\section{Conclusion}
\label{conclusion}

A time- and GPU-efficient unified deep neural network based on 2D CNN for 3D SRR and MAR is proposed. The $1\times1\times2$ down-sampling factors for $\times2$ acceleration and $2\times2\times2$ for $\times4$ acceleration were identified as optimal. TS-RCAN outperformed the 3D networks of DCSRN, mDCSRN, and ReCNN in SRR, and outperformed UNet in MAR, in SSIM/PSNR performance, GPU load, and interference time. In addition, TS-RCAN provided the uncertainty information, which can be used to estimate the quality of the reconstructed images, improving safety in clinical settings.

\setcounter{secnumdepth}{0}
\section{CRediT authorship contribution statement}
\textbf{H.Li:} Conceptualization, investigation, methodology, software, data curation, formal analysis, validation, visualization, original draft writing and editing. \textbf{J.Liu:} Conceptualization, investigation, methodology, software, writing, and editing. \textbf{M.Schell:} Review and editing. \textbf{T.Huang:} Review and editing. \textbf{A.Laure:} Review and editing. \textbf{K.Schregel:} Review and editing. \textbf{J.Jesser:} Review and editing. \textbf{D.F.Vollherbst:} Review and editing. \textbf{M.Bendszus:} Resources, review and editing. \textbf{S.Heiland:} Resources, funding acquisition, conceptualization, investigation, supervision, review, and editing. \textbf{T.Hilgenfeld:} Resources, funding acquisition, conceptualization, investigation, supervision, review, and editing.

\section{Declaration of competing interest}
The authors declare that they have no known competing financial interests or personal relationships that could have appeared to influence the work reported in this paper.

\section{Data availability}
Data will be made available upon request.

\setcounter{secnumdepth}{1}

\end{document}